%% file: paper_dilemma.tex
\documentclass[10pt,conference,a4paper]{IEEEtran}

\usepackage{times}
\usepackage{epsfig}
\usepackage{graphicx}
\usepackage{amsmath}
\usepackage{amssymb}

\usepackage{amsmath}
\usepackage{amsthm}
\newtheorem{definition}{Definition}
\newtheorem{ass}{Assumption}
\newtheorem{lem}{Lemma}
\newtheorem{theo}{Theorem}
\newtheorem{coro}{Corollary}
\usepackage{amssymb}

\usepackage{float}

\newcommand{\s}[1]{\mathcal{#1}}
\newcommand{\class}{\textrm{class}}
\newcommand{\depth}{\textrm{depth}}
\newcommand{\greedy}{\textrm{greedy}}
\newcommand{\explored}{\textrm{explored}}
\newcommand{\best}{\textrm{best}}
\newcommand{\opt}{\textrm{opt}}
\newcommand{\uni}{\textrm{uni}}
\newcommand{\classes}{\textrm{classes}}
\newcommand{\argmin}{\operatornamewithlimits{arg\,min}}
\newcommand{\argmax}{\operatornamewithlimits{arg\,max}}

\usepackage{pifont}
\newcommand{\cmark}{\ding{51}}%
\usepackage{rotating}

\usepackage{multirow} 
\usepackage[ruled, lined, linesnumbered,noend] {algorithm2e}

\usepackage{subfigure}
\usepackage{wrapfig}

\usepackage{subfigure} 

\usepackage[pagebackref=false,breaklinks=true,colorlinks,
bookmarks=false]{hyperref}

\ifCLASSINFOpdf
\else
\fi
\begin{document}
%
\title{Dilemma First Search\\ for Effortless Optimization of NP-hard Problems}

\author{\IEEEauthorblockN{Julien Weissenberg}
\IEEEauthorblockA{CVL, ETH Zurich}
\and
\IEEEauthorblockN{Hayko Riemenschneider}
\IEEEauthorblockA{CVL, ETH Zurich}
\and
\IEEEauthorblockN{Ralf Dragon}
\IEEEauthorblockA{CVL, ETH Zurich}
\and
\IEEEauthorblockN{Luc Van Gool}
\IEEEauthorblockA{CVL, ETH Zurich}
}

\maketitle

\begin{abstract}
To tackle the exponentiality associated with NP-hard problems, two paradigms have been proposed.
First, Branch~\&~Bound, like Dynamic Programming, achieve efficient exact inference but requires extensive information and analysis about the problem at hand.
Second, meta-heuristics are easier to implement but comparatively inefficient.
As a result, a number of problems have been  left unoptimized and plain greedy solutions are used.
We introduce a theoretical framework and propose a powerful yet simple search method called Dilemma First Search (DFS).
DFS exploits the decision heuristic needed for the greedy solution for further optimization.
DFS is useful when it is hard to design efficient exact inference.
We evaluate DFS on two problems: First, the Knapsack problem, for which efficient algorithms exist, serves as a toy example.
Second, Decision Tree inference, where state-of-the-art algorithms rely on the greedy or randomness-based solutions.
We further show that decision trees benefit from optimizations that are performed in a fraction of the iterations required by a random-based search.
\end{abstract}



%
\IEEEpeerreviewmaketitle

\input{paper_dilemma_tree_search_intro}
\input{paper_dilemma_tree_search_related}

\input{paper_dilemma_tree_search_method}

\input{paper_dilemma_tree_search_knapsack}
\input{paper_dilemma_tree_search_decisiontree}
\input{paper_dilemma_tree_search_exp}
\section{Conclusion}\label{sect:con}

In this work, we introduced a theoretical framework for search algorithms, leading to a novel search algorithm called Dilemma First Search~(DFS).
The main idea is to guide the search by revisiting harder-to-make decisions first and hence increase the probability for a better solution.
Dilemma First Search~(DFS) is very efficient and principled, yet also very simple to implement. These are useful properties especially when little information about the problem is known and exact inference is not a viable option.
We demonstrated on two distinct problems that search optimization by DFS leads to an overall performance improvement. 
Future work entails automatically refining the decision heuristic while searching, e.g.~by using backpropagation and learning. 
We also plan to explore much larger data for example from applications in surface fitting~\cite{BodisCVPR2015} or semantic segmentation~\cite{RiemenschneiderECCV2014}.





\noindent
Acknowledgments: This work was supported by the European Research Council project VarCity (273940). www.varcity.eu

{\small
\bibliographystyle{plain}
\bibliography{abbreviation_short,julien2012,hayko2013} 
}

\newpage

\end{document}

%% file: paper_dilemma_tree_search_intro.tex
\section{Introduction}

Search algorithms are of prime importance to deal with the intractable search space associated with NP-hard problems.
Search algorithms are strategies which guide the search to find approximate solutions to problems where 
there is no principled way of inferring an exact one. This is helpful when the search space is too large to be explored in a brute-force fashion. 
However, search algorithms have a challenging nature. Specifically, they
can be (1)~efficient yet problem-specific and hard to design, (2)~relatively inefficient, or (3)~stochastic. 
Efficient (1) search algorithms such as Branch~\&~Bound~(BB) require appropriate problem-specific bound estimates which can take decades to be found~\cite{BruckerDAM1994}. 
In contrast, (2) some, e.g. Genetic Algorithms~(GA), are relatively easy to implement but their optimization is only based on the final outcome.
That is, they sample a candidate solution and test its performance, ignoring some of the information that could be given by a decision heuristic at intermediate steps.
Further, (3) stochastic search algorithms  
suffer from repeatability issues.

In this paper, we present a novel deterministic and efficient search algorithm, denoted as
Dilemma First Search (DFS), to tackle problems which can be framed as a sequence of actions, e.g. decisions or iterations.
We present a framework to retrieve candidate solutions ordered by their likelihood of being optimal.
DFS is derived from this framework, making few assumptions.
We show in the experiments that the assumptions in DFS are reasonable and yield convincing results for two different NP-hard problems.
We consider DFS as ``effortless'' since the only problem-specific requirement is a decision heuristic, which is needed in any case to build a greedy solution. 
DFS is a backjumping algorithm, which uses the evidence at the time of constructing the sequence of actions that some decisions are harder to
make than others (dilemmas). In order to improve an existing solution, we reconsider such dilemmatic actions first. A decision between
several actions is hard to make when they are deemed equally likely to lead to the optimal solution.
Intuitively, solving a search problem consists in finding the sequence of \emph{right} decisions.
Since not all decision combinations can be explored, it is sensible to explore the most likely combinations first.
These can be derived by changing decisions associated with the highest uncertainty first.
Our main contributions are: 


\begin{itemize}
\item a novel search algorithm, Dilemma First Search,
\item a probabilistic theoretical framework for search algorithms in general,

\item optimizations based on DFS for decision tree inference and Knapsack type problems.

\end{itemize}

%% file: paper_dilemma_tree_search_related.tex
\section{Related Work}

\begin{table*} 

\small
\begin{center}
\begin{tabular}{|l|l|c|c|c|c|c|c||c|} \hline
&  & Greedy     & GA       & BB    & LDS           & Tabu          & SA    & DFS       \\ \cline{3-9} \cline{3-9}
& Stochastic            & -     & \cmark        & -             & -     & -     & \cmark        & -     \\ \hline
\parbox[t]{2mm}{\multirow{2}{*}{\rotatebox[origin=c]{90}{Need}}}
& Lower bound  & -     & -     & \cmark                & -     & -     & -             & -     \\
& Decision heuristic    & \cmark        & -     & -             & \cmark     & -     & -             & \cmark        \\ \hline
\parbox[t]{2mm}{\multirow{5}{*}{\rotatebox[origin=c]{90}{Efficiency}}}
& Redundant computation & -     & \cmark        & -             & -     & -     & \cmark                & -     \\
& Space requirement     & small      & little & little & adjustable & adjustable & little & adjustable  \\
& Fast initial solution & \cmark        & -     & -     & \cmark        & \cmark        & \cmark                & \cmark\\
& Reuse computation     & - & - & \cmark                & \cmark        & -     & -             & \cmark\\
& Optimal solution?   & -     & -     & \cmark & \cmark       & \cmark        & -             & \cmark\\ \hline
\end{tabular}
\vspace{1mm}
\caption{Overview of search algorithms considering their requirements and efficiency compared to our
Dilemma First Search.}
\label{tab:overview:searchalgo}
\end{center}
\vspace{-10mm}
\normalsize
\end{table*}


Given the many topics related to search algorithms, we present a brief overview and focus on the two exemplar problems, i.e. the Knapsack and the inference of decision trees.

Some of the most widely used search algorithms are compared in Table~\ref{tab:overview:searchalgo}.
For a complete overview, we refer the reader to~\cite{Korf1999}.
We distinguish between deterministic and stochastic approaches.
First, backtracking~(BT) is the most general deterministic concept. It is guaranteed to find the optimal solution, but suffers from not using heuristics to decide upon the order of the search.
Limited Discrepancy Search~(LDS)~\cite{Harvey1995} is a form of backtracking assuming the best solutions are more likely to be within a local neighborhood from the greedy solution.
Tabu Search~\cite{Glover1989} prevents the search algorithm from re-visiting similar states, thus enforcing diversity.
When an appropriate lower bound can be found, Branch~\&~Bound~(BB) and related Branch~\&~Cut~(BC) prove to be some of the most efficient search algorithms.
We refer the reader to~\cite{Zhang1996} for an overview about BB and for other state-of-the-art methods in this domain to~\cite{Sontag2010,Otten2012}.
Unfortunately, finding a good formulation can be very tedious and it is not unusual that it takes decades~\cite{BruckerDAM1994}.
Stochastic methods such as Genetic Algorithms~(GA) and Simulated Annealing~(SA) use randomness to be able to overcome local minima.
However, their inherent randomness makes repeatability an issue.
%
Monte Carlo tree search (MCTS) algorithms have been successfully used in games, e.g.~Go.
MCTS backpropagates the result of a playout to gather information about the payout distribution of actions in the tree.
The problem is analogous to a multi-armed bandit problem.
Thus, MCTS algorithms, e.g. UCT~\cite{Kocsis2006}, use random sampling and balance exploitation vs. exploration.
This last problem is addressed by many reinforcement learning approaches, such as R-max \cite{Brafman2003}.
The key difference between MCTS and DFS is their application domains.
In MCTS, the probability distribution associated with actions is initially unknown, while DFS uses a \textit{known} probability distribution over the actions.
For all problems where a greedy solution exists, such a distribution is available.


The Knapsack problem has been studied in depth because it is a very simply formulated NP-hard problem. 
\cite{Kellerer2004}~gives a comprehensive overview of the problem.
In decision tree inference, the idea is that a tree
is constructed, wherein each node denotes a
decision which splits up the data to be classified or regressed into
subsets. The leaves of such trees are used for Classification and Regression
problems~\cite{BreimanCART1984}.
%
%
In sequence, the works on ID3~\cite{QuinlanML1986} and C4.5 trees by
Quinlan~\cite{Quinlan1993} focus on optimal attribute
selection. All possible attributes are tested and
 the one giving the maximal information gain (by entropy or Gini
index) is chosen. Yet this formulation uses a greed optimization which is particularly limiting for
high-dimensional data.
The continuation of research led to a combination of weak (pruned
C4.5) classifiers to achieve better generalization. Forests of weak random decision trees were introduced by~\cite{Amit1996}, where the data is subsampled in multiple ways and the parameters are 
randomly selected~\cite{BreimanML2001}. Furthermore, much research has focused on
 how to best combine the ensembles of classifiers~\cite{OzuysalPAMI2010, BreimanML2001}. 
Finally, the optimization of decision trees has been studied by~\cite{Bennett1994, Ketowski2005} and very recently
is an ongoing topic~\cite{Ren2015, Norouzi2015}.
In all, the limits of decision trees can be summarized in their ability to handle high-dimensional data, 
lack of global decision viewing, lack of generalization and extensive requirement of resources in memory and 
computational power. Some of these limits have been overcome in random forests by introducing randomness in the selection of training
data and parameters, as well as using multiple trees. As a result, training yields non-repeatability issues and testing times are higher due to the use of numerous trees.
In contrast, we focus on improving a single tree by optimizing it. 

%

In summary, Dilemma First Search (DFS) proposes an optimized search strategy to explore the search space in a principled way by making full use of the information given by a decision heuristic.

%% file: paper_dilemma_tree_search_method.tex
\section{A probabilistic framework for search algorithms} \label{sect:framework}
We introduce a probabilistic framework for decision making for a sequence of actions.
Further, we show how the exploration of the solution space can be optimally performed.


\subsection{Best sequence of action problem statement}

Formally, the problem $(\s S,\s A, P, E)$ is considered, where $\s
S$ is a finite set of states $S$ and $\s
A$ is a finite set of actions $A$. The state $S = (A_1, ...,A_N)$ is defined by a sequence of $N$ actions.
Each action is associated with a probability $P_{\best}(A_i|S_k)$, as defined in Definition~\ref{def:pbest}. 

\begin{definition} \label{def:pbest}
 Denote by $\hat{S}^*(S_k)$ the optimal candidate solution in a subtree whose root is $S_k$.
$P_{best}(A_i|S_k)$ is the probability that $\hat{S}^*(S_k)$ is contained in the subtree given by taking action $A_i$ from state $S_k$.
\end{definition}

We call a state candidate solution $\hat{S}$ if it solves the problem.
Let $\mathcal{\hat{S}}_{exp}$ and $\mathcal{\hat{S}}_{unexp}$ denote the sets of explored and unexplored candidate solutions, and $\mathcal{S}_{exp}$ and $\mathcal{S}_{unexp}$ the sets of explored and unexplored states.
An algorithm which aims at finding the optimal solution, i.e.~at finding $\hat{S}^* = \operatorname*{arg\,min}_{\hat{S}} E(\hat{S})$ is referred to as a search algorithm.
Note that $P_{best}(A_i|S_k)$ and the energy function $E(\hat{S})$ are problem-specific.
Due to the combinatorial nature, $(\s S,\s A, \s P, E)$ is an NP-hard problem.
We can represent this problem by a tree structure, where nodes are states and edges are actions:
$S_k \xrightarrow{A_k \in \s A_k} S_{k+1}$, where $\s A_k$ is the set of actions at state $S_k$.
In this tree, each path from the root to a leaf corresponds to a candidate solution $\hat{S}$.
Please note that selecting the maximum $P_{\best}(A_k|S_k)$ at each state gives the greedy solution.
An example of a search tree is given in Fig.~\ref{fig:dfsidea}.

Given an unknown, arbitrary  and limited amount of computational resources, we want to maximize the probability that we retrieve the optimal solution.
Hence, we want to evaluate candidate solutions in order of their likelihood of being optimal.

\subsection{Fully explored search tree}
When the tree is fully explored, the probability of each candidate solution $\hat{S}$ being the optimal $\hat{S}^*$ is given by $P_{\opt}(\hat{S})$:

\begin{equation} \label{eq:candidateProba}
 P_{\opt}(\hat{S}) = \prod_{k=0}^{\depth(\hat{S})} P_{\best}(A_k|S_k).
\end{equation}

Please note, the tree needs to be fully explored to guarantee the candidate solutions are evaluated in order of decreasing probability of being the optimal: when the tree is partially explored,
the probabilities $P_{\best}(A_k|S_k)$ are not all known.

\subsection{Partially explored search tree}

Exploring the full tree is time and memory consuming and can often not be done in practice.
Therefore, we require an anytime algorithm. An anytime algorithm is one which is able to return intermediate solutions of increasing quality, i.e. which can be stopped at ``any time''~\cite{Hendler1992}.
The idea is to estimate which state to explore first in a partially explored tree. In order to do so, we have to be able to compute the expected probability of partially explored candidate solutions $\hat{S}$.
For this, we make the following assumptions.



\begin{ass} \label{ass:proba}
$P_{\best}(A|S_{unexp}) = P_{\uni}$
i.e. the expected values of the probabilities of actions are uniformly distributed in unexplored parts of the tree. $P_{\uni} = \frac{1}{b}$, where $b$ is the tree branching factor.
\end{ass}

\begin{ass} \label{ass:depth}
The expected depth $d$ of the tree is uniform and known.
\end{ass}

\begin{ass} \label{ass:branchingfactor}
The branching factor $b$ of the tree is constant and known.
\end{ass}

In the following, we first explain how to calculate candidate solution probabilities in the tree.
We show that the greedy solution is most likely the best candidate solution (base case).
In addition, given a partially explored tree where the first $n$ most likely candidate solutions have been found, the $n+1^{th}$ most likely candidate solution can be inferred (inductive step).

We first calculate the expected probability of each candidate solution consisting of explored and unexplored states following As.~\ref{ass:proba}, \ref{ass:depth} and \ref{ass:branchingfactor}:


\begin{equation} \label{eq:popt}
 P_{\opt}(\hat{S}) = \prod_{k=1}^{\explored} P_{\best}(A_k|S_k) \cdot \prod_{k=\explored+1}^{d} P_{\uni}
\end{equation}
where $A_k$ maximizes $P_{\best}$ at $S_k$ and $P_{\uni}$ is the uniform probability distribution.




\begin{lem} \label{lem:bounderies}
  $ P_{\uni} \le P_{\greedy}(A_k|S_k) \le 1$ \\
  where $P_{\uni}$ is the uniform probability distribution and $P_{\greedy}(A_k|S_k) = \operatorname*{arg\,max}_{A} P_{\best}(A_{i,k}|S_k)$, $A_{i,k}$ being the $i^th$ action which can be taken at state $S_k$.
\end{lem}


\begin{theo}[Base case] \label{theo:greedy}
When the tree is unexplored, the greedy solution finds $\operatorname*{arg\,max}_{\hat{S}} P_{\opt}(\hat{S})$ when exploring at most depth $d$ nodes.
\end{theo}

\begin{figure*} [t]
\centering
\includegraphics[width=0.32\linewidth]{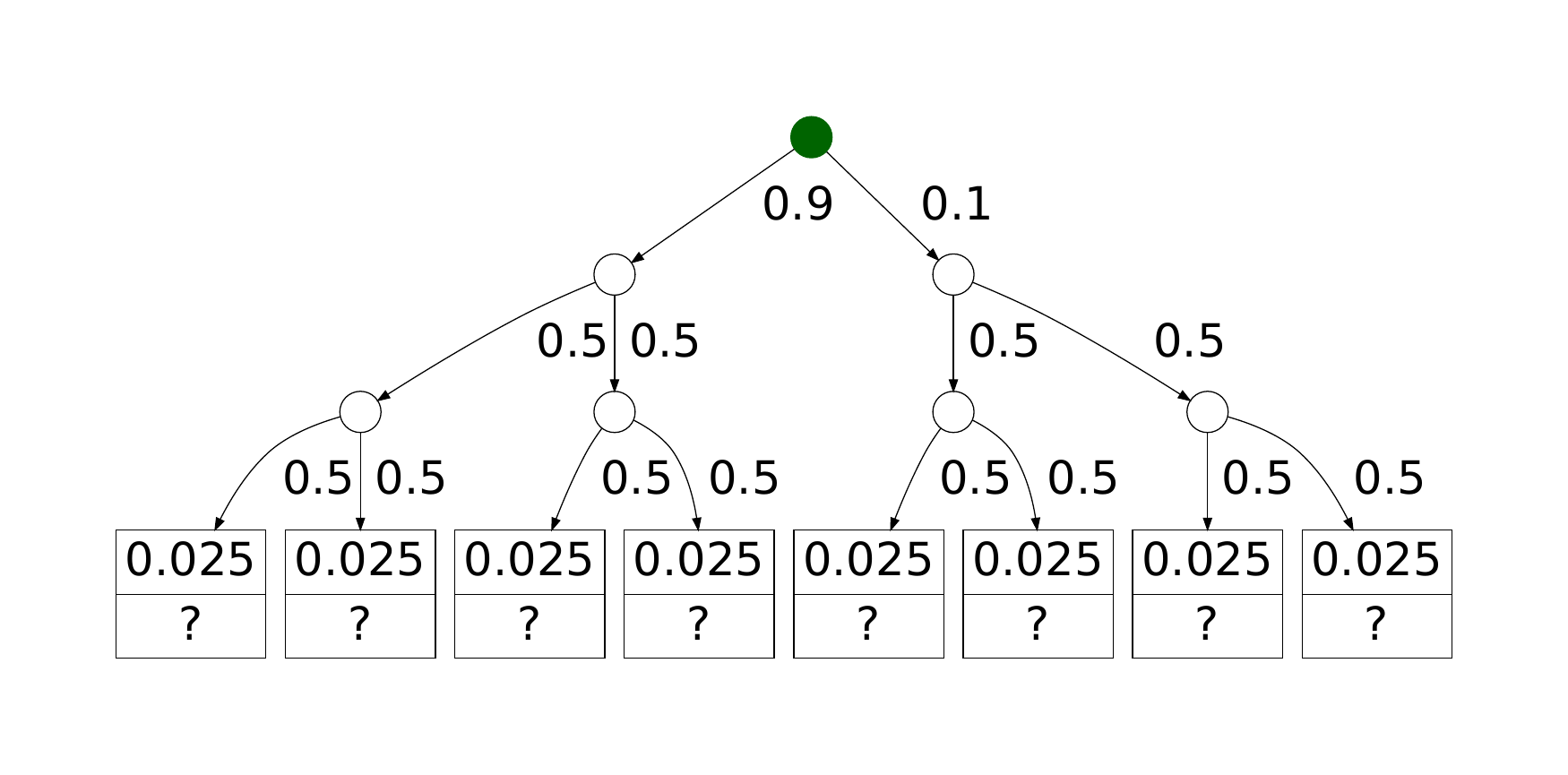}
\includegraphics[width=0.32\linewidth]{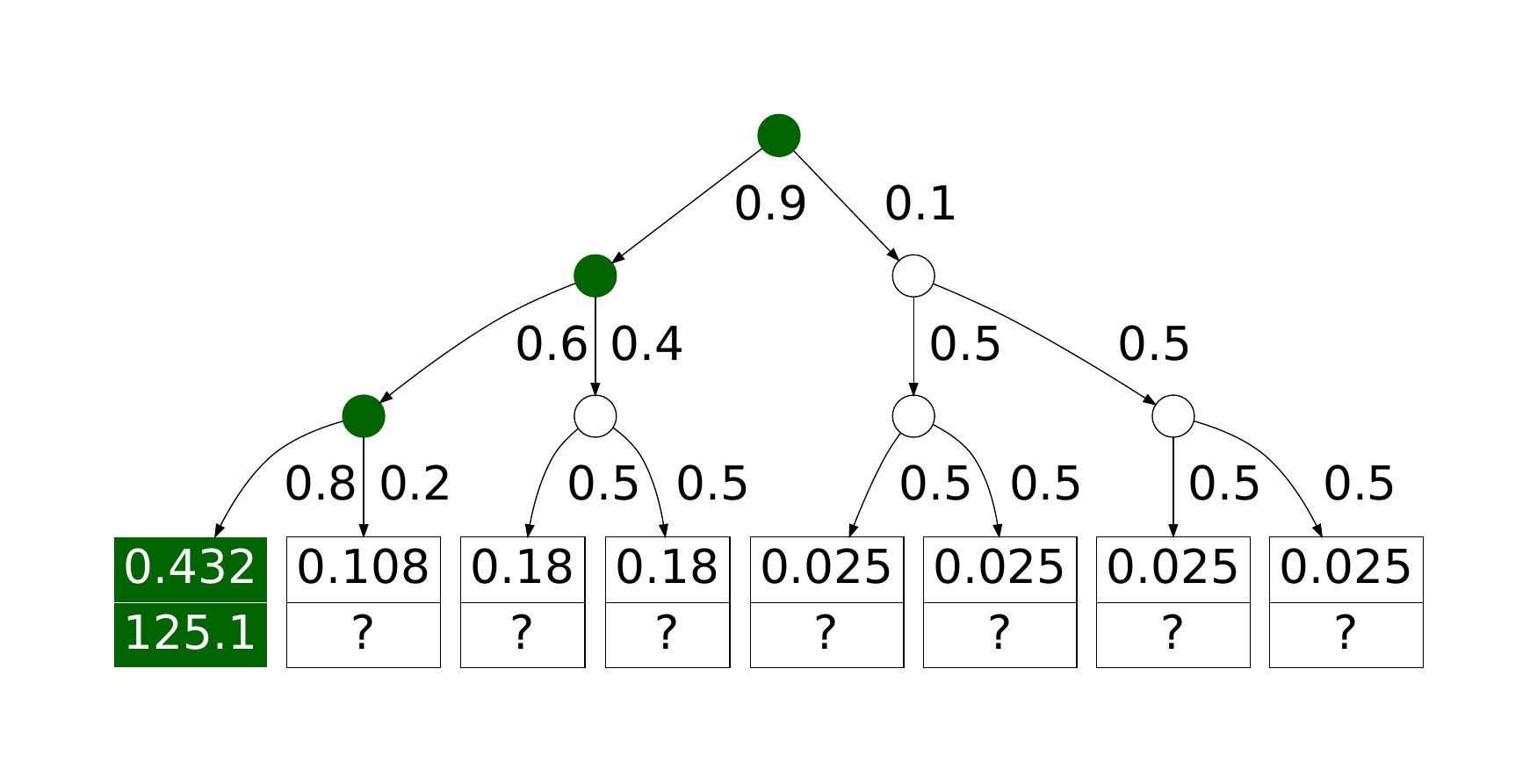}
\includegraphics[width=0.32\linewidth]{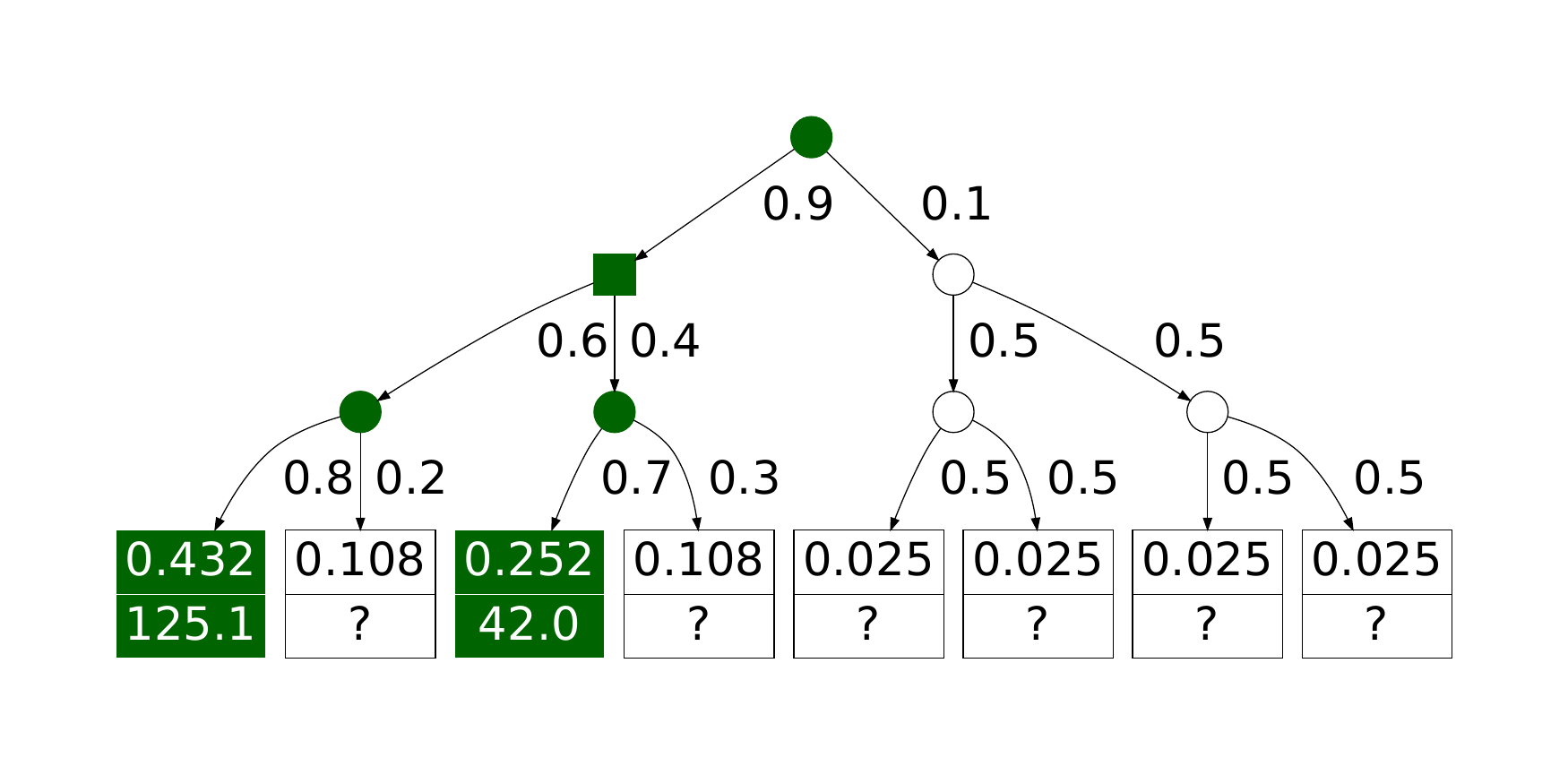}
\caption{Example of Dilemma First Search. Each node corresponds to a state $S$, each edge is an action associated with a probability $P(A | S)$.
The leaf nodes represent the set of candidate solution $\hat{\s S}$ and are labelled with the probability that they are the optimal solution $\hat{S}^*$, along with their energy $E(\hat{S})$ when explored.
Nodes are left blank when they are unexplored and coloured in dark green otherwise.
The dilemma node is indicated by a square.
In the first iteration (left and centre), the tree is initialized and a greedy solution is built. Then (right), the most dilemmatic node is selected and a new candidate solution is grown.}
\label{fig:dfsidea}
\end{figure*}



We now want to find the next candidate solution that maximizes $P_{\opt}(\hat{S})$.


\begin{theo} [Induction] \label{theo:next}
  Given the set of $n^{th}$ most likely candidate solutions have been explored, the $n+1^{th}$ most likely candidate solution are inferred by exploring~at~most~d~more~nodes.
  
\end{theo}

\begin{coro} \label{coro:bestisalpha}
 The average number of candidate solutions to be evaluated in order to find $\hat{S}^*$  is smallest when $\alpha$ is chosen according to
 $\operatorname*{arg\,min}_{\alpha} (L_{\opt}(\hat{S_{n}}) - L_{\alpha})$, which can be approximated by $\operatorname*{arg\,min}_{\alpha} ( L_{\best}(S_{\alpha}) - L'_{\best}(S_\alpha) - \depth(\alpha) \cdot const.)$ (See Appendix in supp. mat. for the derivations).
\end{coro}

\subsection{Search algorithm strategies} \label{subsect:strategies}

We now informally examine the implications of the results of the previous section and relate them to existing search algorithms.
Note that probabilities can be approximated by a normalized decision heuristic.
If the probability distribution has a low entropy, the search is expected to be more efficient.
Conversely, if probabilities are distributed more uniformly, the search is expected to require more iterations since less information can be leveraged. 
First, if the computational resources are limited to computing a single candidate solution, then computing the greedy solution yields the best chances of obtaining the optimal solution.
Moreover, from Theorem~\ref{theo:next} and Eq.~\eqref{eq:popt}, altering high-quality sets of solutions is expected to yield higher than average quality candidate solutions. 
In particular, Simulated Annealing~(SA) and Genetic Algorithms~(GA) respectively use the best candidate or set of candidates so far to build new candidate solutions.
It is also sensible to substitute the greedy solution with the second best choice when backtracking, as in Limited Discrepancy Search (LDS).
However, LDS is sub-optimal as it performs back-tracking but does not select the optimal node to explore given the information at hand.
Finally, we note that Branch~\&~Bound~(BB) assumes more information is available, namely an estimate of the lower and upper bounds.
These are not always easily available and have to be manually designed for each problem.
In the following section, we propose an algorithm based on the theoretical derivations presented so far.

\section{Dilemma First Search} \label{sect:dilemma}

An efficient search algorithm based on the probabilistic search framework consists of three parts: (1)~selecting
a dilemmatic state $S_\alpha = (A_1,... ,A_n)$; (2)~changing its action $A_n$ to $A_n'$; (3)~growing a complete candidate solution starting from the partial solution of the selected state $S_\alpha$. 
An example illustration of these steps for improving a decision tree is shown in Figure~\ref{fig:dfsidea}.
Before we detail our method and its properties, we give an intuition of the method.

\textbf{Intuition:} \label{sect:approach}
The proposed method is similar to the way humans improve sequences of
decisions. Suppose you would like to find the fastest path to the top of a mountain without a map. At every
intersection, a decision has to be made. The steepness of the path is
a good indicator and can be used as a decision guide.  A first
route can be obtained by taking the path with the most appropriate
steepness (greedy search).  To improve an existing route (candidate solution),
it is sensible to first revisit the path intersections (states) where you
were most hesitating.
At dilemmatic intersections the selected path was not considerably better than the second best.
In other words, going up the $2^{nd}$-best option is likely to yield a new, competitive route ($2^{nd}$-best candidate solution).





  \begin{algorithm}
    \small
    \caption{Dilemma First Search (DFS)}
    \KwData{$(\s S,\s A, \s P, E)$}
    \KwResult{$\hat{S}^*$}
    \Begin{
      $\mathcal{\hat{S}}_{exp} \gets \emptyset$; $\s Q \gets \emptyset$; $\s A_S \gets \emptyset \quad \forall \s S$; 
      $S_\alpha \gets ()$\; 
      \While{termination criterion not met} {
        {\footnotesize  \tcc{greedy candidate search}}
        $S \gets S_\alpha$\;
        \While{$S$ is not a candidate solution}{
          $A \gets \argmax_{A \in \s A \backslash \s A_S} P(A | S)$\;
          $\s A_S \gets \s A_S \cup \{A\}$\; 
          $S \gets (S, A)$\;
          $\s Q \gets \s Q \cup \{S\}$\; 
        }
        $\mathcal{\hat{S}}_{exp} \gets \mathcal{\hat{S}}_{exp} \cup \{S\}$\;

        { \footnotesize \tcc{select dilemma state $S_\alpha$}}
        $S_\alpha \gets \argmax_{S \in \s Q} \delta(S)$\;
        $\s Q \gets \s Q \backslash \{S_\alpha\}$\;
      }
      $\hat{S}^* \gets \argmin_{\hat{S} \in \mathcal{\hat{S}}_{exp} } E(\hat{S})$\;
    }
    \normalsize

  \end{algorithm}
  

\textbf{Algorithm:} \label{sect:algo} DFS optimizes a sequence of actions $\hat{S}$ to minimize an energy function $E(\hat{S})$, by revisiting 
actions which were most dilemmatic first.
When a decision is revisited, it will be altered to the next-best choice indicated 
by $P(A | S)$ and the subsequent actions from that state are chosen in a 
greedy fashion until a candidate solution is reached.

%

As displayed in Algorithm~1, 
in the first iteration, a greedy solution~(l.~5--9) is built.
Simultaneously, each explored state $S$ is filed into a dilemma
queue~$\s Q$~(l.~9).
When a candidate solution is reached, it is added to the set of candidate solutions $\hat{\s S}$~(l.~10).
In the following iterations, the state with the largest dilemma estimator $\delta(S_\alpha)$ is selected and removed from the queue (l.~11-12).
The dilemma estimator $\delta(S_\alpha)$ is derived as:
\begin{align} \label{eq:dilemmaEstimator}
        \delta(S_\alpha) = 1/({L_{\best}(S_\alpha) - L_{\best}'(S_\alpha)})
\end{align}

Eq.~\eqref{eq:dilemmaEstimator} follows from Cor.~\ref{coro:bestisalpha}, 
where const. is assumed to be close to zero.
This assumption is valid if the alternative path is as likely to yield the optimal solution as the original path from which it is derived from.





\textbf{Anytime} means that DFS returns intermediate solutions of increasing quality (l. 6). The more computational resources are available, the more likely the solution will be optimal.

\textbf{Convergence} is guaranteed since DFS eventually explores the whole search tree and therefore converges to the optimal solution. Please note, the ordering of exploration is important.
DFS retrieves most promising candidate solutions first.

The two following properties show how DFS relates to the probabilistic framework given in Sect.~\ref{sect:framework}.
\textbf{First iteration} of DFS yields the greedy solution, in accordance with Theorem~\ref{theo:greedy}.\\ 
\textbf{Subsequent iterations} create new competitive candidate solutions by selecting the most dilemmatic state (see Eq.~\eqref{eq:dilemmaEstimator}), switching to the next best action and greedily growing until a candidate solution has been reached.

\textbf{DFS vs Random state selection (RSS)}: RSS consists in selecting a state at random from the set of explored states, regardless of their dilemma estimator.
In other words, RSS only differs from DFS in that it randomizes the order of the dilemma queue.
RSS requires more iterations on average to find~$\hat{S}^*$ than selecting the dilemma state based on the optimal~$S_{\alpha}$ (see Corollary~\ref{coro:bestisalpha}).
Therefore, we investigate RSS as a comparison to DFS in the experimental section.

%% file: paper_dilemma_tree_search_knapsack.tex
\section{Knapsack problem} \label{sect:knapsack}

In this section, we explain how DFS can be used to search for solutions of the 0-1 Knapsack problem efficiently.
The 0-1 Knapsack problem is formulated as follows.
Given $n$ items $x_i = (w_i, v_i) \in X$, where $w_i$ is the weight of item $x_i$ and $v_i$ its value,
we would like to select a subset of items $X_s \in X$, such that the total profit $\sum_{{v_j}\in X_s} v_j$
is maximized and the total weight $\sum_{{w_j}\in X_s} w_j < W$ does not exceed the knapsack capacity $W$. 
For this classical NP-hard problem, a pseudo-polynomial time dynamic programming solution~\cite{Torth1980} and a Branch and Bound solution~\cite{Kolesar1967} exist.
This section serves as a toy example to explain how DFS is applied.

\textbf{Greedy approach} \label{sect:knapsackgreedy}
First, we define a decision heuristic which will help choosing the items.
From any current state $S$, the greedy approach iteratively adds an item which is not in the knapsack, while fitting. We use the heuristic~\cite{MartelloKnapsack1990}:

\begin{equation} \label{eq:ksheuristic}
 f(A_i|S) = \frac{v_i}{w_i} \text{~if~} w(S) + w_i \leq W\text{, 0 otherwise}
\end{equation}

\noindent where $A_i$ corresponds to adding item $x_i$ to the Knapsack.
The greedy approach is to always choose the decision with the maximum heuristic score, until no more item can be added.

\textbf{Dilemma First Search for Knapsack}
In addition to the heuristic $f(A_i|S)$ (see Eq.~\eqref{eq:ksheuristic}), we specify an energy function $E(\hat{S}) = v(\hat{S})$ (where $v(\hat{S})$ is the knapsack value, i.e. the sum of the values of items in the Knapsack), 
and a dilemma estimator $\delta(S)$.
The dilemma estimator $\delta(S)$ is calculated following Eq.~\eqref{eq:dilemmaEstimator} where $L_{\best}(S_\alpha)$
is estimated as $f(A_1|S)$ and $L_{\best}'(S_\alpha)$ as $f(A_2|S)$, $f(A_i|S)$ being defined as in Eq.~\eqref{eq:ksheuristic}.


%% file: paper_dilemma_tree_search_decisiontree.tex
\section{Decision Tree Optimization} \label{sect:dt}

In this section, we detail how Dilemma First Search (DFS) helps in decision tree
learning. First, we give a brief overview of the ID3 algorithm in our notation. Then,
we explain how to optimize the initial greedy  solution.

The decision tree to be created is used to classify data vectors $\vec x = (x_1, \ldots,
x_n)$. Without loss of generality, let each $x_j$ with $j=1\ldots n$ be a categorical value
from the finite set of attributes~$\s X_j$.
Each node in the tree partitions the space according to one of the attributes $x_j \in \s X_j$.
Finally, a leaf node signals the classification result.
In order to establish a decision tree, the set of training data $\s T = \{(\vec x;\textrm{class}(\vec x)),\dots \}$ is used, each component $\vec x$ being labelled with $\class(\vec x) \in \classes(\s T)$.

\textbf{Greedy approach}
We re-use the heuristic from the ID3 algorithm~\cite{QuinlanML1986}.
Based on the current state $S=(A_1,\ldots,A_{i-1})$ which corresponds to a partially
established tree with $(i-1)$ nodes, the action $A_{i}$ is selecting the
attribute for the decision of the next $i^{\textrm{th}}$ node. To
determine the attribute~ $j := A_i$, the information
gain is used as heuristics

\begin{equation}
  \label{eq:id3heuristic}
  f(A_i | S) = H(\s T) - \sum_{x \in \s X_j} \frac{|\s T_j(x)|}{|\s T|}
  H(\s T_j(x))\,,
\end{equation}

where $\s T$ is the training data available for the node,
$\s T_j(x)$ is the subset of $\s T$ with the $j^{th}$ attribute being equal to $x$,
and $H$ the class entropy defined by the proportion $p$ of occurrences of
class $c$ in $\s T$:

\begin{equation}
  \label{eq:entropy}
  H(\s T) = - \sum_{c \in \classes(\s T)} p(c| \s T) \cdot \log_2 p(c| \s T) \,.
\end{equation}
 
After $j$ is determined, the child nodes with all $x \in \s X_j$ are
explored using Eq.~\eqref{eq:id3heuristic} with the filtered training data $\s T' = \s T_j(x)$.
The algorithm finishes when all the attributes have been used, a maximum depth is reached, or when the leaf nodes have a small enough entropy.

\textbf{Dilemma First Search for ID3}
Having defined the ID3 heuristic $f(A | S)$ in
Eq.~\eqref{eq:id3heuristic}, we now specify the energy function $E(\hat{S})$ and
the dilemma estimator $\delta(S | \s A_S)$ according to
Section~\ref{sect:algo}. To prevent overfitting, we define the energy $E$
as the number of mis-classifications over a validation set $\s V$
which is disjoint to the training data as
\begin{equation}
  \label{eq:classenergy}
  E(\hat{S}) = {|\{\vec x \in \s V | d_S(\vec x) \ne \class(\vec x)\}|}\,,
\end{equation}

where $d_S(\vec x)$ is the classification of the data vector $\vec x$
carried out with the tree specified by the candidate solution~$S$.
The dilemma estimator $\delta(S)$ is calculated following Eq.~\eqref{eq:dilemmaEstimator} where $L_{\best}(S_\alpha)$ and $L_{\best}'(S_\alpha)$
are estimated as $f(A_i|S)$, $f(A_i|S)$ being defined as in Eq.~\eqref{eq:id3heuristic}.


%% file: paper_dilemma_tree_search_exp.tex
\section{Experimental Evaluation} \label{sect:exp}

\begin{figure*}[ht]
\begin{tabular}{ccc}
\includegraphics[width=0.3\linewidth,clip,trim=3.6mm 00mm 2.3mm 0mm]{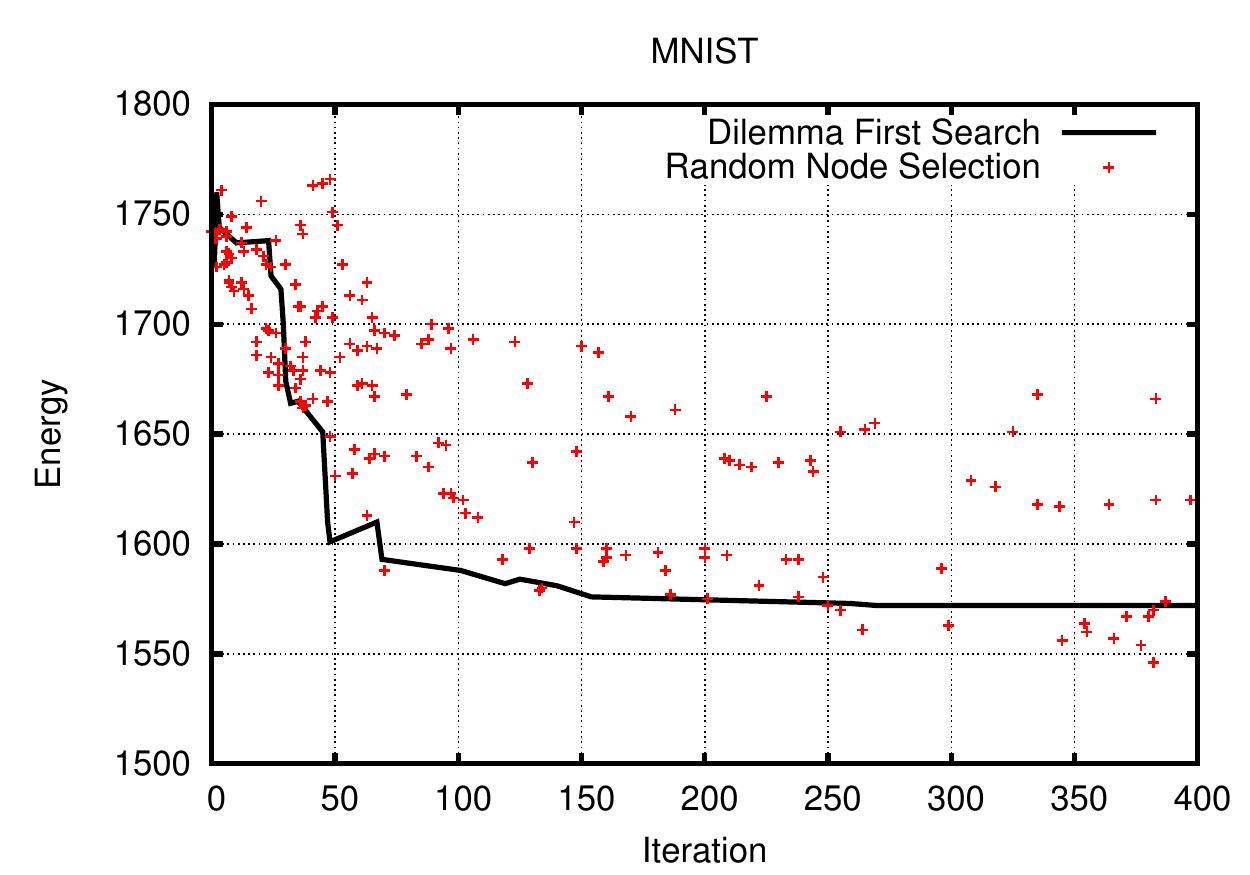} 
& \includegraphics[width=0.3\linewidth,clip,trim=3.6mm 00mm 2.3mm 0mm]{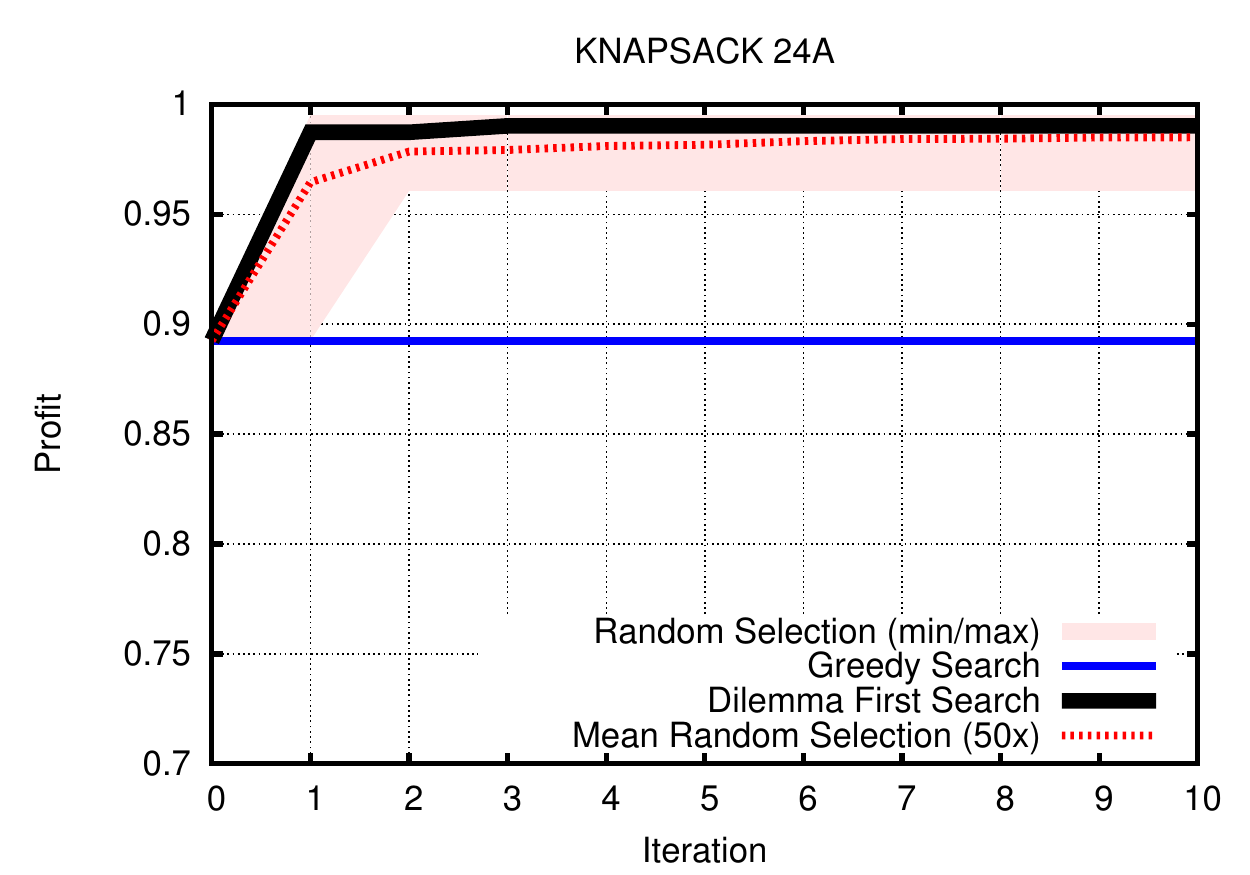}
& \includegraphics[width=0.3\linewidth,clip,trim=3.6mm 00mm 2.3mm 0mm]{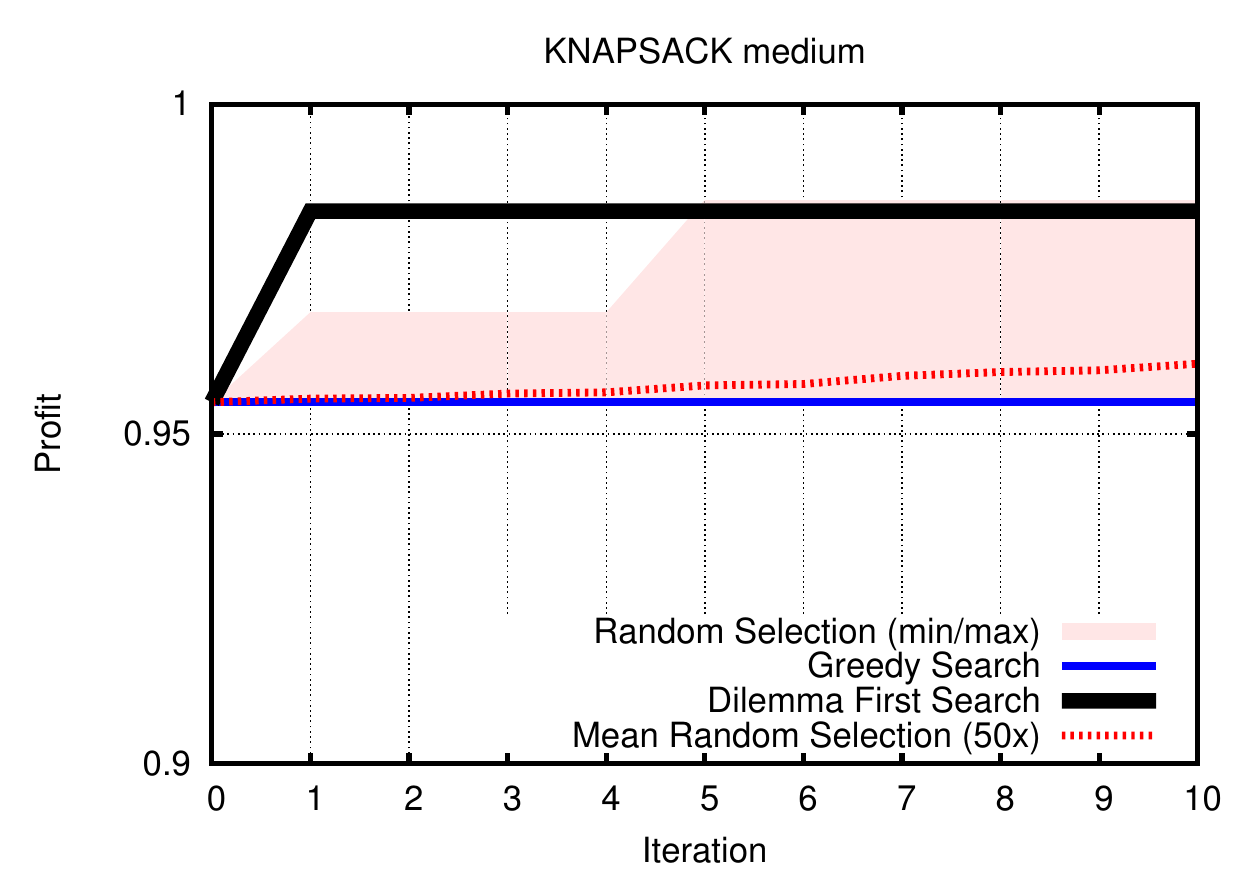} \\

\includegraphics[width=0.3\linewidth,clip,trim=3.6mm 00mm 2.3mm 0mm]{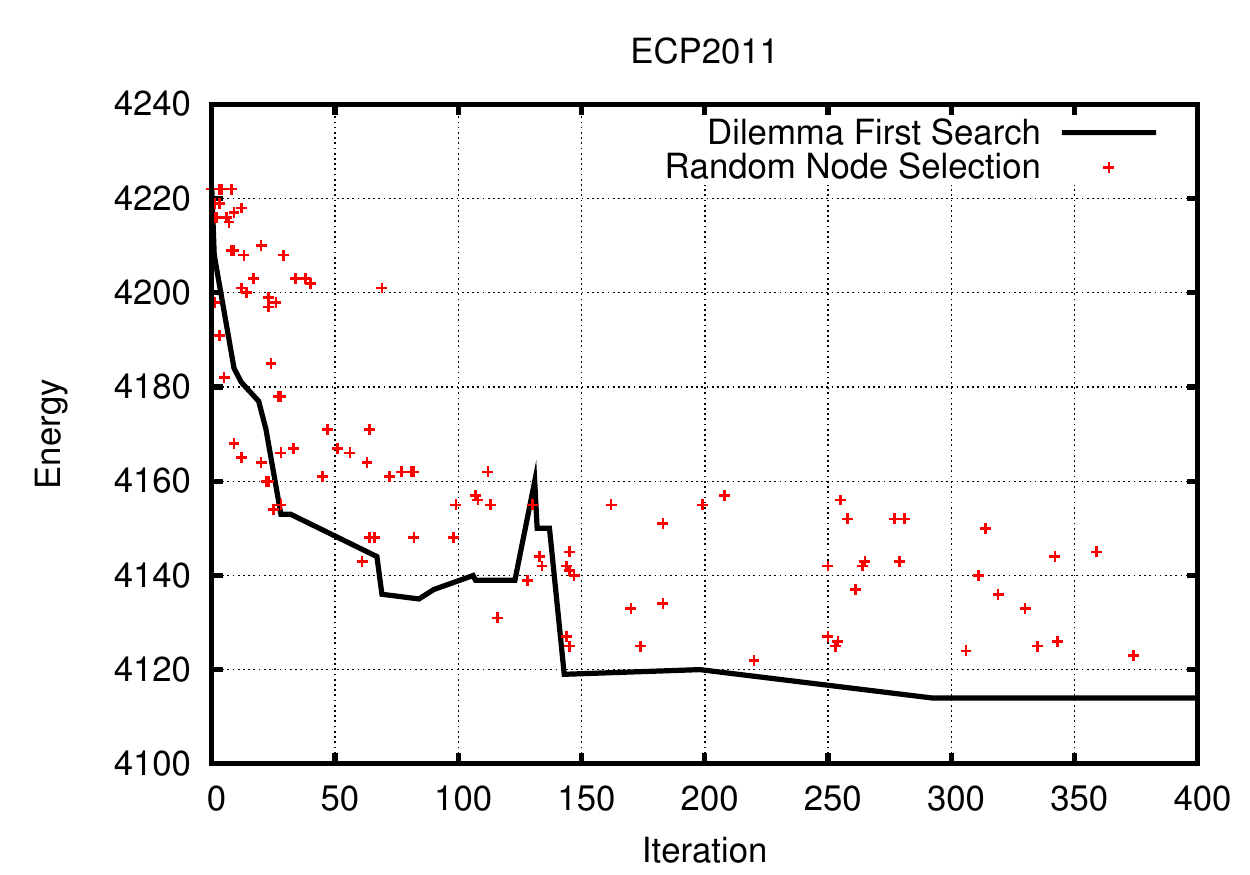}
& \includegraphics[width=0.3\linewidth,clip,trim=3.6mm 00mm 2.3mm 0mm]{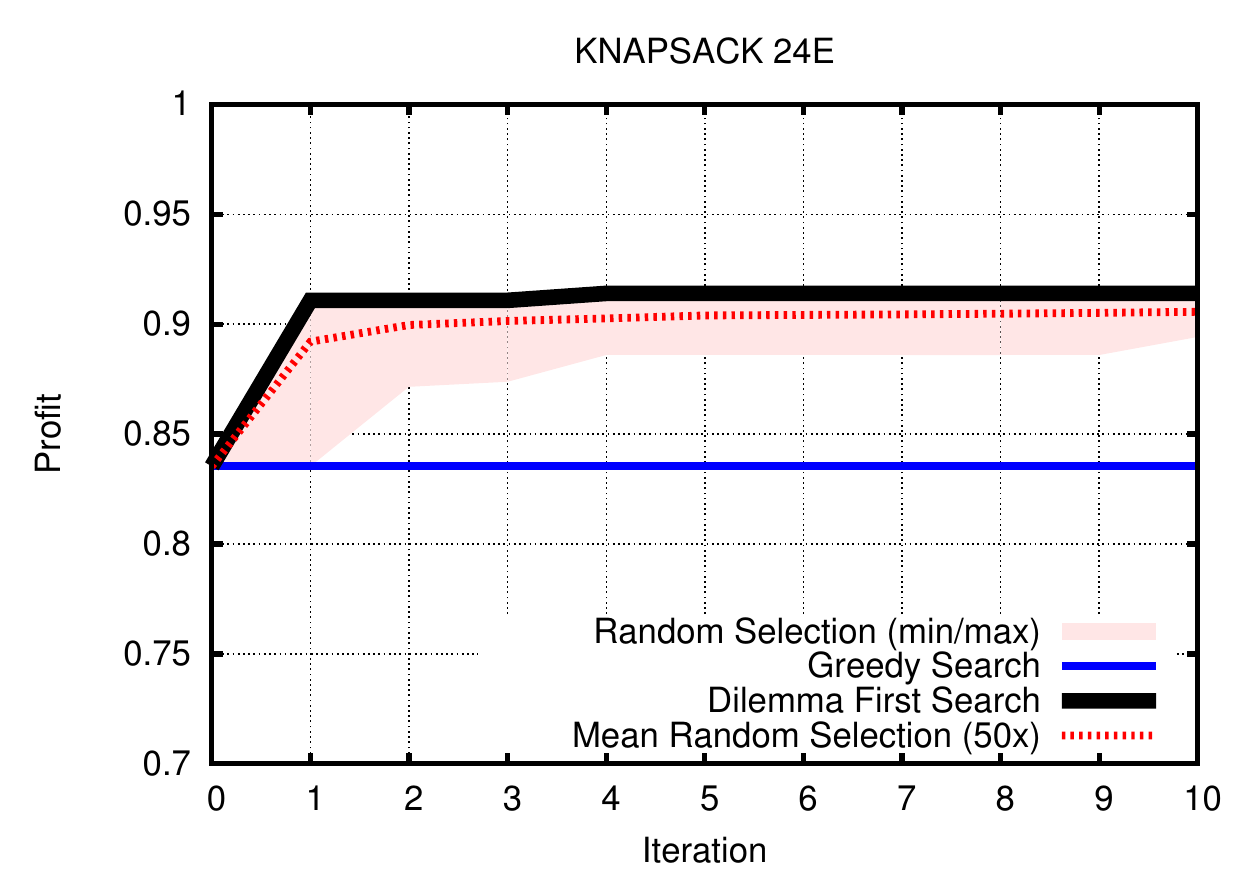}
& \includegraphics[width=0.3\linewidth,clip,trim=3.6mm 00mm 2.3mm 0mm]{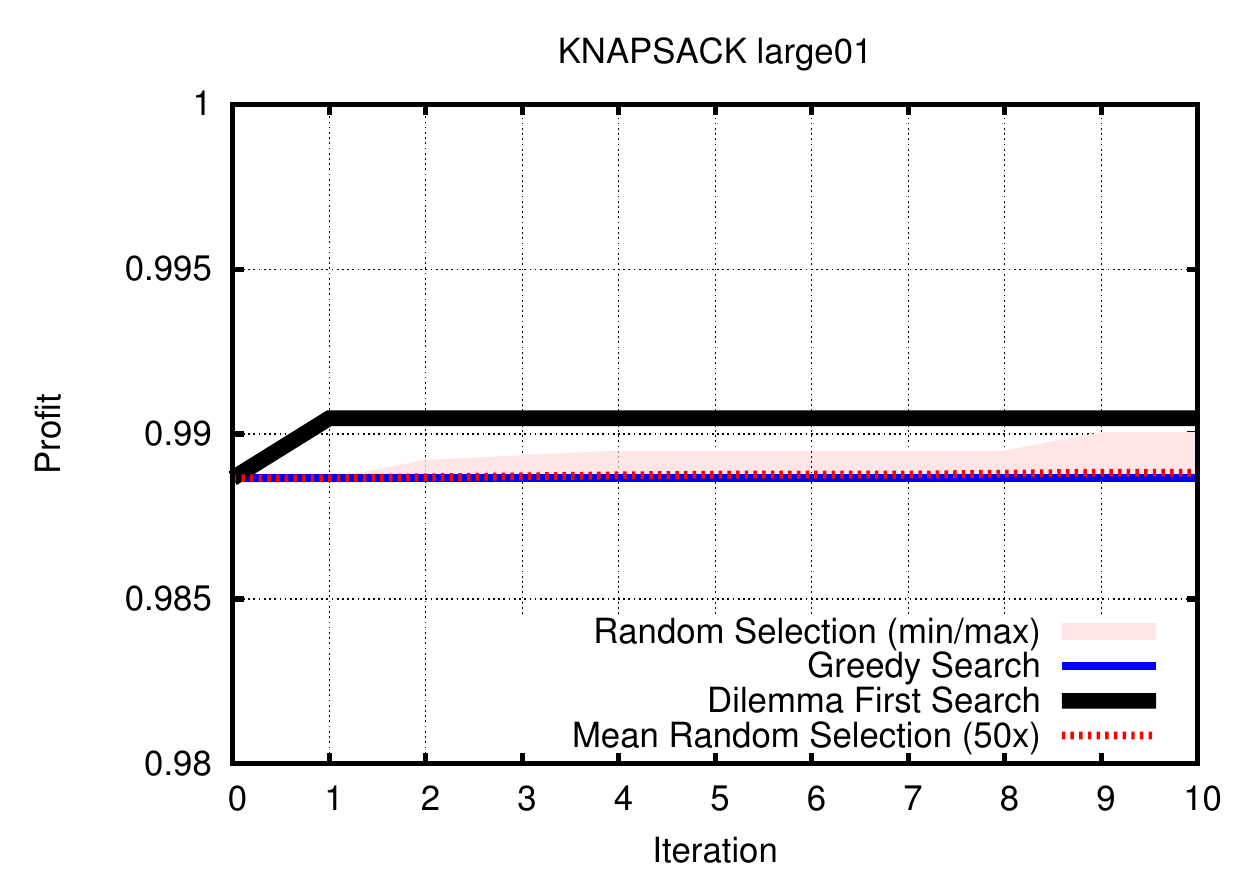} \\
Decision tree & Knapsack & Knapsack (large)
\end{tabular}
\caption{
Comparison on Decision tree (left) and Knapsack optimizations (right) between DFS (black) and Random Node Selection (RNS) (red, as a point cloud for Decision trees and averaged over 50 trials for Knapsack). The 0th iteration corresponds to the greedy solution. RNS picks a state at random and explores the next best candidate solution according to a given heuristic. DFS picks the most dilemmatic node and explores the next best candidate solution according to a given heuristic. Picking the most dilemmatic state results in almost systematic improvement in convergence speed over RNS.
}
\label{fig:results}
\end{figure*}

We considered four standard datasets for the 0-1 Knapsack problem, as well as two diverse datasets for decision tree optimization.
For both Knapsack and Decision Tree, each iteration takes less than 100 ms (unoptimized code).
Note that the runtime depends on the times for computing (1) the energy of a candidate solution, (2) the decision heuristics, (3) the sorted dilemma queue.
(1+2) are the same than for the greedy algorithm while (3) can be efficiently achieved in $\mathcal{O}(\sqrt{log(n)})$ time with the Fusion tree algorithm by \cite{Fredman1990}.
As discussed above, we note that it is relevant to compare DFS against random state selection. 
If DFS is an efficient informed search, we should notice that DFS outperforms random state selection on average (from Corollary~\ref{coro:bestisalpha}).

\subsection{Knapsack Results}

As a first experiment, we analyze the 0-1 Knapsack problem.
The greedy solution selects the items in order of best value per weight as described in Section~\ref{sect:knapsackgreedy}. This provides an good initial solution. See Figure~\ref{fig:results} for results where the greedy solution is the $0^{\textrm{th}}$ iteration. 
We evaluate on four standard datasets~\cite{KreherCombinatorialAlgos1998,MartelloKnapsack1990} of various sizes (24A/E with 24, medium with 100, and large01 with 10000 items). 
In all cases, DFS is able to improve over the greedy solution in few iterations, although the search space is extremely large. 
Further we consistently outperform random state selection (averaged over 50 random runs).
The random state selection picks a random state and regrows a greedy solution from that state. Therefore, the only difference to DFS occurs in line 15 of the DFS algorithm.
Not only is DFS better than random state selection on average for all datasets, it also consistently outperforms the best random state selection result.




\vspace{-2mm}
\subsection{Decision Tree Results}

As a second experiment, we optimize the construction of decision trees for classification.
The greedy solution is the one of the ID3 algorithm, which greedily selects the most discriminating attribute for classification. 
We evaluate our proposed search optimization and its effects on two different datasets. 
The MNIST dataset~\cite{LeCun1998} is a text character recognition dataset, which consists of size-normalized handwritten digits to be classified.
It consists of a 60,000 instance training set and of a 10,000 instance test set.
For this dataset, the raw binary image is simply the input to the decision tree. 
Next, we evaluate DFS optimization on the semantic segmentation dataset ECP2011~\cite{TeboulIJCV2010}.
It comprises 104 images and seven semantic classes, where the goal is to classify each pixel
into one of the semantic classes.
The RGB values of each pixel are quantized into 10 bins each, and each bin combination constitutes a binary attribute. 
As the protocol suggests, we split the dataset into two disjoint sets
containing 70\% data for training and
validation, and 30\% for testing. To analyze the performance of DFS, we evaluate the energy of the
testing data according to Eq.~\eqref{eq:classenergy}. 
For both decision trees, we enforce a maximal depth of eight to prevent the tree from overfitting.
This reduction in depth also results in improved classification time, as the entire decision tree classifier is smaller.
Please note, a limited depth makes state selection even more crucial. 


Figure~\ref{fig:results} shows results for the two datasets, where the greedy solution is the $0^{\textrm{th}}$ iteration in all graphs.
Starting from a high mis-classification rate, our method improves the classification rate more efficiently than a random state selection -- in ECP2011 by $2.6\%$ and in MNIST by $9.7\%$.
The random state selection achieved a reduction by only $1.9\%$ and $7.6\%$ on average, respectively. 
Furthermore, DFS is almost always converging faster than any of the random state selection runs.
For this problem, the depth of the search tree being fixed, we evaluated over a range for the constant in Cor.~\ref{coro:bestisalpha} 
according to lower/upper bounds of the estimated probabilities.
There was no significant change in the results, confirming that a preference to reuse large parts of the solution is not required.